\documentclass[aps,twocolumn,prd,showpacs,showkeys,preprintnumbers,superscriptaddress,nobibnotes,floatfix,longbibliography,nofootinbib]{revtex4-1}

\usepackage{amsmath}
\usepackage{amsfonts}
\usepackage{amssymb}
\usepackage{mathrsfs}
\usepackage{graphicx}
\usepackage{color}
\usepackage{hyperref}
\usepackage{multirow}
\usepackage{slashed}
\usepackage{epsf}
\usepackage{color}

\begin{document}

\title{Natural Dark Matter in Generalized No-scale $\mathcal{F}$-$SU(5)$ at the HL-LHC and LUX-ZEPLIN Experiment}

\author{Tianjun Li}
\email{tli@itp.ac.cn}
\affiliation{School of Physics, Henan Normal University, Xinxiang 453007, P. R. China}

\author{James A. Maxin}
\email{JamesMaxin@MissouriState.edu}
\affiliation{Missouri State University, West Plains, MO 65775, USA}

\author{Dimitri V. Nanopoulos}
\email{dimitri@physics.tamu.edu}
\affiliation{Academy of Athens, Division of Natural Sciences, 28 Panepistimiou Avenue, Athens 10679, Greece.}
\affiliation{George P. and Cynthia W. Mitchell Institute for Fundamental Physics and Astronomy,
	Texas A$\&$M University, College Station, TX 77843, USA}


\begin{abstract}

Is a supersymmetric bulk region with sub-200 GeV neutralinos and right-handed sleptons still feasible? Initially the bulk was the discovery target of the supersymmetry (SUSY) pioneers in the 1980s, though reality crashed those ambitions over the subsequent years and decades, presumptively excluding these light neutralinos and sleptons, inviting finely tuned dark matter scenarios to satisfy emerging high-precision relic density measurements. The answer to this important question is $yes$, if we consider the GUT model $\mathcal{F}$-$SU(5)$, namely, flipped $SU$(5) supplemented with a string derived TeV-scale extra $10+\overline{10}$ vector-like multiplet and charged vector-like singlet $(XE,XE^c)$, dubbed flippons. We pair $\mathcal{F}$-$SU(5)$ with effective supernatural supersymmetry (SUSY), known as generalized No-Scale Supergravity (SUGRA). This merging of $\mathcal{F}$-$SU(5)$ with generalized No-Scale SUGRA solves the SUSY electroweak fine-tuning problem and includes natural dark matter (NDM) that meets the latest 5$\sigma$ Planck satellite observations through annihilation only. The NDM in this model is imminently under test by the LUX-ZEPLIN 1000-day experiment, thus prospective observed events by LUX-ZEPLIN in 2026-27 will compel confirmation by the high-luminosity LHC (HL-LHC) energizing in 2030. Therefore, we present the parameter space observable at both LUX-ZEPLIN and the HL-LHC that is consistent with all presently operating experiments, including the LHC Run 3 gluino and squark constraints. Our results predict sub-200 GeV neutralinos and right-handed sleptons in the bulk that are not beyond detection, though more massive electroweakinos greater than 600 GeV are also allowed.

\end{abstract}

\maketitle


\section{Introduction}

The theoretical beauty of supersymmetry (SUSY) has long been demonstrated to resolve several dilemmas beyond the scope of the Standard Model (SM). A few of the more significant historical successes of the supersymmetric SM (SSM) with R parity are (1) gauge coupling unification, as observed at LEP~\cite{gaugeunification}; (2) the lightest supersymmetric particle (LSP) neutralino as a viable dark matter (DM) candidate~\cite{Goldberg:1983nd,Ellis:1983ew}; and (3) electroweak (EW) gauge symmetry can be radiatively broken by the large top quark Yukawa coupling~\cite{Ellis:1982wr}. Perhaps the most far-reaching attribute is that superstring theory provides a framework from which grand unified theories (GUTs)~\cite{Georgi:1974sy,Pati:1974yy,Mohapatra:1974hk,Fritzsch:1974nn,Georgi:1974my,Buras:1977yy} can be constructed, implying gauge coupling unification. Although, these ordinary GUTs need a Higgs adjoint representation to break the GUT symmetry, which is not possible in string theory. In contrast, flipped $SU(5)$~\cite{Barr:1981qv,Derendinger:1983aj,Antoniadis:1987dx} naturally includes antisymmetric $\mathbf{(10, \overline{10})}$ GUT-scale Higgs representations without adjoints such that the GUT symmetry is broken uniquely down to the Standard Model in string theory. Hence, SUSY can be regarded as a fundamental link between string theory and low-energy phenomenology, opening an observation window to new beyond the SM physics.

The search for statistically signifcant SUSY events at the LHC has thus far produced a null set, though in reality when we reckon with a conceivable string landscape, the probing of the SUSY parameter space is only in its infancy. The lack of a substantiated discovery has elevated low mass bounds on the gluino, first-two generation squarks, stop, and sbottom to around 2.25~TeV, 1.9~TeV, 1.25~TeV, and 1.5~TeV, respectively~\cite{ATLAS-SUSY-Search, Aad:2020sgw, Aad:2019pfy, CMS-SUSY-Search-I, CMS-SUSY-Search-II}, indicating that the SUSY Electroweak (EW) fine-tuning (EWFT) problem may need to be addressed. There have been numerous positive solutions to the EWFT problem proposed~\cite{Dimopoulos:1995mi,Cohen:1996vb,Kitano:2005wc,Kitano:2006gv,LeCompte:2011cn,LeCompte:2011fh,Fan:2011yu,Kribs:2012gx,Baer:2012mv,Baer:2012cf,Drees:2015aeo,Ding:2015epa,Baer:2015rja,Batell:2015fma,Fan:2014axa}, including Super-Natural SUSY~\cite{Leggett:2014hha,Du:2015una, Li:2015dil} by considering No-Scale Supergravity (SUGRA)~\cite{Cremmer:1983bf} and SUSY breaking soft terms in M-theory on $S^1/Z_2$~\cite{Li:1997ii,Li:1997sk,Li:1998rn}. Moreover, it was shown in No-Scale SUGRA that that high-energy fine-tuning measure defined by Ellis-Enqvist-Nanopoulos-Zwirner~\cite{Ellis:1986yg} and Barbieri-Giudice~\cite{Barbieri:1987fn} (EENZ-BG) is naturally of order one for both light and heavy spectra, and that Effective Super-Natural SUSY~\cite{Ding:2015epa} can be generalized from Super-Natural SUSY~\cite{Leggett:2014hha,Du:2015una, Li:2015dil}.

Accompanying the EWFT problem, demonstrating consistency with dark matter relic density measurements for dominantly bino LSPs persists as a further daunting predicament. Solutions that are commonly applied include: (i) light sfermions, defined as the bulk region; (ii) LSP neutralino mass is around half of the $Z$ boson, SM Higgs, $CP$-even Higgs $H_{0}$, or $CP$-odd Higgs $A_{0}$ masses, referred to as the $Z$/Higgs funnel or $Z$/Higgs resonance; (iii) sfermion masses are close to the LSP neutralino mass, defined as coannihilation; or (iv) LSP neutralino has sufficient wino or Higgsino component to substantially increase the annihilation cross section, known as the mixing or well-tempered scenario. In consideration of these four options, the light sfermion bulk region is clearly the most natural solution, while the other three options seem to engage in fine tuning the bino mass in order to satisfy relic density measurements. We shall present model predictions in this work for present and forthcoming experiments that resolve the EWFT problem and prevent a finely tuned bino mass.

Constructing a bulk region in contemporary times is non-trivial, given the relentless upward march of LHC mass limits on gluinos and squarks over the prior 16 years. Observation of the bulk was the opening salvo in the 1980s in the battle for SUSY discovery~\cite{Ellis:1983ew} given its naturally light neutralinos and right-handed sleptons. However, the dearth of persuasive SUSY signals at LEP, Tevatron, and LHC have rendered the bulk an afterthought in the present day. Despite this four decade gloomy history of the bulk, it was recently analyzed in both generalized No-Scale ${\cal F}$-$SU(5)$~\cite{Yin:2023jhs} and the phenomenological minimal SSM (pMSSM)~\cite{Yin:2024bwv} and found to be alive and well, but stealthily concealed, though not beyond the reach of the LHC Run 3 (LHC3) and High Luminosity LHC (HL-LHC), and further down the road, the Future Circular Collider (FCC-ee)~\cite{FCC:2018byv,FCC:2018evy} at CERN and Circular Electron Positron Collider (CEPC)~\cite{CEPCStudyGroup:2018ghi}. Possible observation is not limited to only the LHC and circular colliders, but the bulk is also imminently under probe at the 1000-day LUX-ZEPLIN experiment~\cite{LZ:2018qzl,LZ:2022lsv,LZ:2024zvo}. The primary goal in this work is to offer an update of the ongoing LUX-ZEPLIN intrusion into the generalized No-Scale ${\cal F}$-$SU(5)$ model space and predictions for the highly anticipated powering up of the HL-LHC beams around the year 2030.

We would be remiss in not highlighting the exclusive model aspects of ${\cal F}$-$SU(5)$, {\it i.e.}, the flipped $SU(5)\times U(1)_X$ GUT model~\cite{Barr:1981qv,Derendinger:1983aj,Antoniadis:1987dx} with extra TeV-scale vector-like particles~\cite{Jiang:2006hf} constructed systematically in local F-theory model building~\cite{Jiang:2009zza,Jiang:2009za}. Models of this type have also be realized in free fermionic string constructions~\cite{Lopez:1992kg}. In the instance of a light bino LSP, Super-Natural SUSY via No-Scale SUGRA cannot resolve the SUSY EWFT problem given that the bino and wino masses are correlated. However, we proposed generalized No-Scale SUGRA in order to realize Effective Super-Natural SUSY. Up to the LHC3, SUSY searches have revealed that all sfermions other than right-handed sleptons must be heavy. Therefore, we can only consider light right-handed sleptons if we wish to identify the bulk region for dark matter. In order to do this, an interaction between sfermions and the LSP must be inspected to classify it as either annihilation or coannihilation. This can be accomplished via the mass difference $\mathcal{R}_{\phi}\equiv (m_{\phi}-m_{\widetilde{\chi}_{1}^{0}})/m_{\widetilde{\chi}_{1}^{0}}$~\cite{Yin:2023jhs,Yin:2024bwv}, with $\phi$ representing $\widetilde{\tau}_{1}$ (light stau) or $\widetilde{e}_R$ (light selectron). Note that the absolute mass difference is not as important as this ratio. To ensure the observed dark matter density is computed from only traditional annihilations, and not coannihilations or resonances, etc., it has already been demonstrated~\cite{Yin:2023jhs,Yin:2024bwv}that $\mathcal{R}_{\phi} \gtrsim 10\%$ is a conservative benchmark to delineate the bulk region.

\begin{table*}[ht]
\scriptsize
\begin{center}
\begin{tabular}{c|c|c|c|c|c|c||c|c|c|c|c|c|c|c|c|c|c|c|c|c|c} \hline
$	M_{1X}$&$	M_5	$&$	M_0	$&$	M_{Hu/Hd}	$&$	A_0	$&$	M_V	$&$	tan\beta	$&$	\Omega_{\chi} h^2	$&$	m_{\widetilde{\chi}^0_1}	$&$	m_{\widetilde{\chi}^0_2}	$&$	m_{\widetilde{\chi}^{\pm}_1}	$&$	m_{\widetilde{\tau}^{\pm}_1}	$&$	m_{\widetilde{e}_L/\widetilde{\mu}_L}	$&$	m_{\widetilde{e}_R/\widetilde{\mu}_R}	$&$	m_{\widetilde{t}_1}	$&$	m_{\widetilde{u}_R}	$&$	m_{\widetilde{g}}	$&$	m_h	$&$	m_{A^0}	$&$	\sigma_{SI}	$&$	\mathcal{R}_{\widetilde{\tau}_1}	$&$	\mathcal{R}_{\widetilde{e}_R}	$	\\ \hline \hline
$	340	$&$	1778	$&$	9.83	$&$	96.54	$&$	0	$&$	5912	$&$	4.880	$&$	0.1194	$&$	95	$&$	773	$&$	773	$&$	105	$&$	1740	$&$	147	$&$	1977	$&$	3205	$&$	2302	$&$	125.46	$&$	3221	$&$	1.4	$&$	10.8\%	$&$	54.6\%	$	\\ \hline
$	310	$&$	1870	$&$	16.02	$&$	256.80	$&$	0	$&$	5543	$&$	4.244	$&$	0.1185	$&$	90	$&$	812	$&$	812	$&$	101	$&$	1831	$&$	138	$&$	2066	$&$	3369	$&$	2411	$&$	125.27	$&$	3453	$&$	1.5	$&$	12.0\%	$&$	53.1\%	$	\\ \hline
$	351	$&$	1960	$&$	2.98	$&$	8.88	$&$	0	$&$	6646	$&$	4.542	$&$	0.1251	$&$	101	$&$	858	$&$	858	$&$	111	$&$	1903	$&$	152	$&$	2168	$&$	3501	$&$	2529	$&$	125.17	$&$	3540	$&$	1.3	$&$	10.3\%	$&$	50.1\%	$	\\ \hline
$	352	$&$	2059	$&$	5.55	$&$	30.79	$&$	0	$&$	6201	$&$	4.341	$&$	0.1248	$&$	102	$&$	901	$&$	901	$&$	112	$&$	1999	$&$	153	$&$	2276	$&$	3676	$&$	2645	$&$	125.62	$&$	3737	$&$	1.2	$&$	10.0\%	$&$	50.0\%	$	\\ \hline
$	308	$&$	2192	$&$	1.91	$&$	3.66	$&$	0	$&$	5754	$&$	3.550	$&$	0.1157	$&$	95	$&$	958	$&$	958	$&$	105	$&$	2125	$&$	139	$&$	2399	$&$	3906	$&$	2799	$&$	124.31	$&$	4115	$&$	1.5	$&$	10.8\%	$&$	46.6\%	$	\\ \hline \hline
$	320	$&$	1826	$&$	9.47	$&$	89.63	$&$	0	$&$	4822	$&$	4.508	$&$	0.1055	$&$	91	$&$	788	$&$	788	$&$	100	$&$	1798	$&$	140	$&$	2031	$&$	3310	$&$	2351	$&$	126.53	$&$	3365	$&$	1.4	$&$	10.3\%	$&$	54.3\%	$	\\ \hline
$	317	$&$	1911	$&$	9.89	$&$	97.83	$&$	0	$&$	5896	$&$	4.277	$&$	0.1070	$&$	92	$&$	832	$&$	832	$&$	102	$&$	1865	$&$	140	$&$	2110	$&$	3431	$&$	2463	$&$	125.17	$&$	3508	$&$	1.4	$&$	10.3\%	$&$	51.7\%	$	\\ \hline
$	312	$&$	1999	$&$	13.26	$&$	175.88	$&$	0	$&$	6129	$&$	4.039	$&$	0.1066	$&$	93	$&$	873	$&$	873	$&$	102	$&$	1945	$&$	139	$&$	2201	$&$	3576	$&$	2572	$&$	124.69	$&$	3684	$&$	1.5	$&$	10.2\%	$&$	50.2\%	$	\\ \hline
$	314	$&$	2067	$&$	5.25	$&$	27.55	$&$	0	$&$	4810	$&$	3.890	$&$	0.1069	$&$	94	$&$	896	$&$	896	$&$	103	$&$	2023	$&$	140	$&$	2280	$&$	3719	$&$	2640	$&$	126.58	$&$	3861	$&$	1.4	$&$	10.1\%	$&$	49.8\%	$	\\ \hline
$	299	$&$	2186	$&$	16.53	$&$	273.30	$&$	0	$&$	5027	$&$	3.563	$&$	0.1031	$&$	93	$&$	950	$&$	950	$&$	102	$&$	2130	$&$	137	$&$	2402	$&$	3914	$&$	2784	$&$	125.41	$&$	4122	$&$	1.5	$&$	10.0\%	$&$	48.2\%	$	\\ \hline \hline
$	324	$&$	1794	$&$	8.79	$&$	77.23	$&$	0	$&$	4974	$&$	4.220	$&$	0.1586	$&$	91	$&$	775	$&$	775	$&$	108	$&$	1766	$&$	141	$&$	1988	$&$	3252	$&$	2313	$&$	125.36	$&$	3340	$&$	1.7	$&$	18.6\%	$&$	54.6\%	$	\\ \hline
$	358	$&$	1909	$&$	2.10	$&$	4.39	$&$	0	$&$	4166	$&$	3.606	$&$	0.2081	$&$	100	$&$	821	$&$	821	$&$	129	$&$	1887	$&$	153	$&$	2102	$&$	3470	$&$	2443	$&$	125.85	$&$	3670	$&$	1.9	$&$	29.2\%	$&$	54.0\%	$	\\ \hline
$	314	$&$	1985	$&$	14.60	$&$	213.12	$&$	0	$&$	5123	$&$	3.759	$&$	0.1537	$&$	92	$&$	861	$&$	861	$&$	108	$&$	1944	$&$	140	$&$	2181	$&$	3574	$&$	2546	$&$	125.17	$&$	3740	$&$	1.6	$&$	17.0\%	$&$	51.4\%	$	\\ \hline
$	371	$&$	2090	$&$	1.68	$&$	2.82	$&$	0	$&$	6432	$&$	4.344	$&$	0.1676	$&$	107	$&$	916	$&$	916	$&$	120	$&$	2025	$&$	160	$&$	2313	$&$	3723	$&$	2684	$&$	125.33	$&$	3778	$&$	1.2	$&$	12.6\%	$&$	49.8\%	$	\\ \hline
$	346	$&$	2183	$&$	4.63	$&$	21.45	$&$	0	$&$	2345	$&$	2.214	$&$	0.2122	$&$	100	$&$	926	$&$	926	$&$	136	$&$	2191	$&$	150	$&$	2289	$&$	4020	$&$	2756	$&$	126.20	$&$	5091	$&$	2.6	$&$	36.7\%	$&$	50.5\%	$	\\ \hline \hline
\end{tabular}
\end{center}
    \caption{Fifteen generalized No-Scale $\mathcal{F}$-$SU(5)$ benchmark points that reside in the bulk, separated into three groups for undersaturated, saturated, and oversaturated relic density. All masses are in GeV. The numerical values for $\sigma_{SI}$ are in units of $10^{-11}$~pb, planted precisely within the region bisected by the LUX-ZEPLIN 1000-day experiment ``Median 3$\sigma$ discovery potential'' demarcation line, as illustrated in the right-side panel of FIG.~\ref{atlas_luxzeplin}. The light Higgs boson mass $m_h$ includes the maximal vector-like flippon Yukawa coupling contribution at 1-loop, in addition to the 2-loop coupling to the heavy light stops via the top Yukawa coupling.}
     \label{table}
\end{table*}

{\bf The $\mathcal{F}$-$SU(5)$ Model~--}~A comprehensive review of ${\cal F}$-$SU(5)$ can be studied in Refs.~\cite{Li:2010ws,Li:2011ab,Li:2021cte}, and references therein. For more information on the No-Scale flipped $SU(5)$ string derived inflationary model, cosmology, and gravitational waves, see Ref.~\cite{Ellis:2019jha}. Here we merely highlight a small number of the model's strongest attributes. Three families of SM fermions that exist in the No-Scale flipped $SU(5)$ model with quantum numbers under the $SU(5)\times U(1)_{X}$ gauge group are
$F_i={\mathbf{(10, 1)}},~ {\bar f}_i={\mathbf{(\bar 5, -3)}},~{\bar l}_i={\mathbf{(1, 5)}}$, where $i=1, 2, 3$.  Introduction of two pairs of Higgs fields $H={\mathbf{(10, 1)}},~{\overline{H}}={\mathbf{({\overline{10}}, -1)}}, ~h={\mathbf{(5, -2)}},~{\overline h}={\mathbf{({\bar {5}}, 2)}}$ breaks the GUT and electroweak gauge symmetries. Introduction of the vector-like particles (dubbed ``flippons'') $XF={\mathbf{(10, 1)}}~,~{\overline{XF}}={\mathbf{(\overline{10}, 1)}}~,~Xl={\mathbf{(1, -5)}}~,~{\overline{Xl}}={\mathbf{(1, 5)}}$, forming complete flipped $SU(5)\times U(1)_X$ multiplets, achieves true string-scale gauge coupling unification~\cite{Jiang:2006hf, Jiang:2009zza, Jiang:2009zza}. The merging of flipped $SU(5)$ with the extra vector-like particle multiplets derived from local F-Theory model building is alluded to as $\mathcal{F}$-$SU(5)$. The $SU(5)\times U(1)_X$ unification is lifted to the string scale around $10^{17}$~GeV (correlating to the mass scales $M_5$ and $M_{1X}$), adjacent to the Planck scale, by renormalization group equation (RGE) $\beta$ coefficients that undergo a shift due to the vector-like multiplets. Thereafter, near the traditional MSSM GUT scale around $10^{16}$~GeV, we have a second stage unification $SU(3)_C\times SU(2)_L$ (defined as mass scale $M_{32}$). As a result, the dual unification structure engenders a flat $SU(3)$ RGE running (resulting from vanishing $b3$ coefficients enforced by the vector-like multiplets) from the GUT scale to the TeV scale, at which point all vector-like particles decouple, which defines the scale $M_V$, a free model parameter. A benefit of light $M_V$ mass scales ($M_V \lesssim 10$~TeV) is a larger vector-like particle Yukawa coupling, which permits a larger contribution to the light Higgs boson mass~\cite{Li:2011ab}.

Generalization of No-Scale SUGRA is required to achieve a light bino while simultaneously evading large LHC constraints on the gluino mass, all while preserving naturalness conditions. To produce a large gluino mass, we float the $SU(5)$ gaugino mass $M_5$ from 1200~GeV to 5000~GeV at the $SU(5)\times U(1)_X$ ultimate unification scale, which is the string scale. Concurrently, we also float the $U(1)_X$ gaugino mass $M_{1X}$ from 100~GeV to 600~GeV to produce a light bino. Note our assumption is that the universal SUSY breaking soft mass $M_0$ and trilinear soft term $A_0$ are smaller than about 1\% of $M_5$ since No-Scale SUGRA is obtained at tree level and can be violated at one loop. Moreover, we fluctuate $\tan\beta$ from 2 to 65, and the vector-like particle mass scale $M_V$ from 1 TeV to 10 TeV. It was shown that a SUSY breaking scenario is natural if the agregate collection of fundamental parameters that have large EENZ-BG fine-tuning measures are correlated, this scenaro defined as Effective Super-Natural SUSY~\cite{Ding:2015epa}. In generalized No-Scale SUGRA, fine-tuning measures for the SUSY breaking soft terms $M_{1X}$, $M_0$, and $A_0$ are small, though $M_5$ could have a large fine-tuning factor. In this case, there is only a small deviation from Super-Natural SUSY, therefore, generalized No-Scale SUGRA is approximately Super-Natural SUSY and thus natural. The reason for this is that in Effective Super-Natural SUSY, only a single fundamental parameter can have a large EENZ-BG fine-tuning measure, furnishing the simplest case for Effective Super-Natural SUSY.

\begin{figure*}
    \centering
    \begin{tabular}{c c}
        \includegraphics[width=1\textwidth, trim = {0 10cm 0 0}, clip]{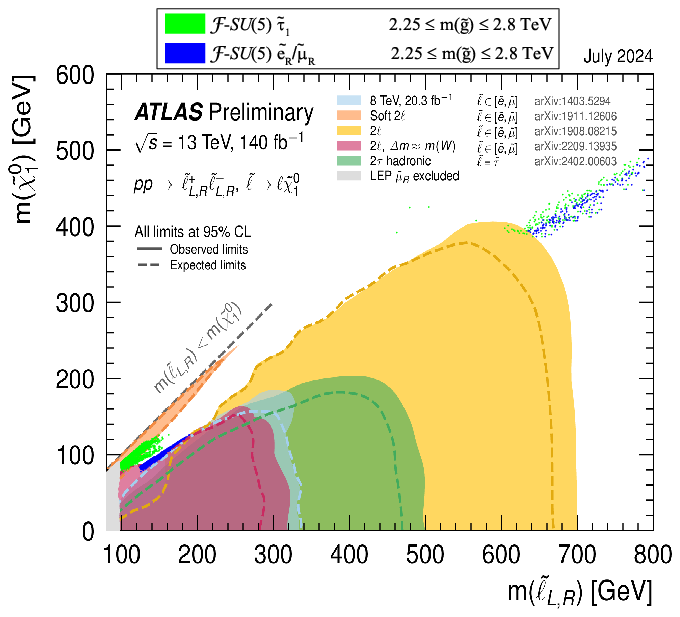} & \hspace*{-10.0cm}\includegraphics[width=1\textwidth, trim = {0 10cm 0 0}, clip] {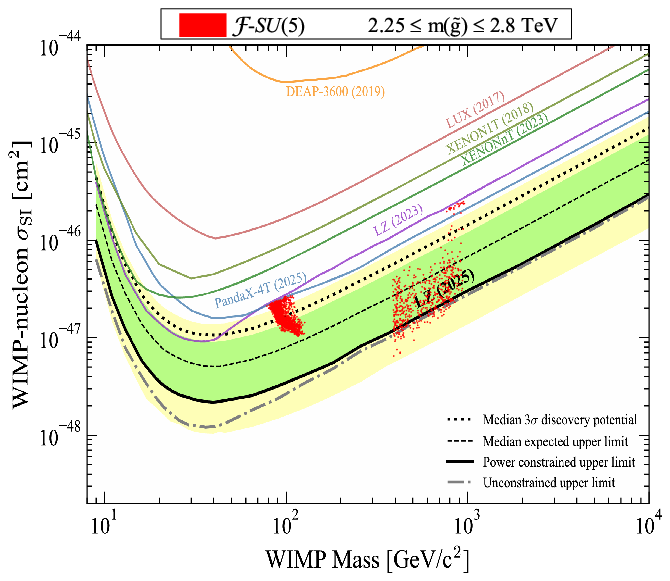} \\
        \includegraphics[width=1\textwidth, trim = {0 10cm 0 0}, clip]{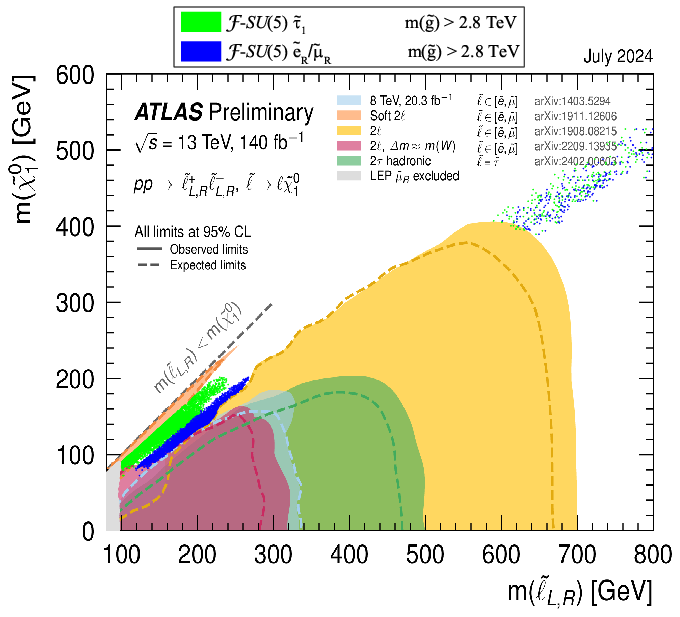} & \hspace*{-10.0cm}\includegraphics[width=1\textwidth, trim = {0 10cm 0 0}, clip] {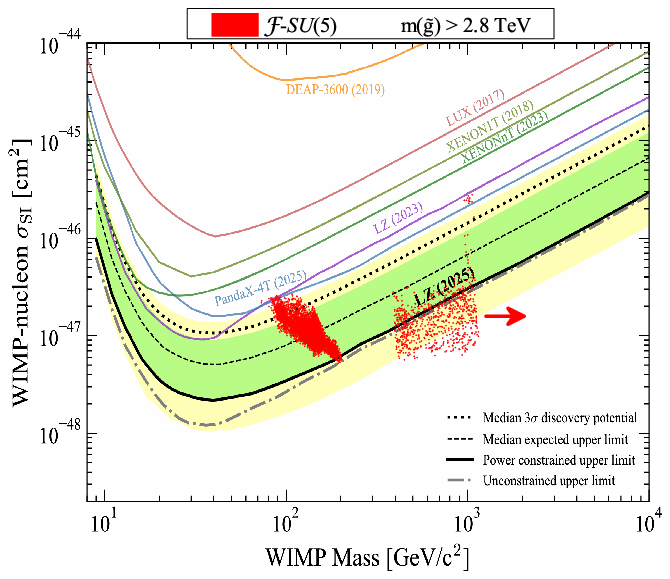}
     \end{tabular}
    \caption{Generalized No-Scale $\mathcal{F}$-$SU(5)$ model space superimposed upon the summary plot of ATLAS SUSY searches for electroweak production of sleptons~\cite{ATLAS:2023xco,ATLAS:2014zve,ATLAS:2019lng,ATLAS:2019lff,ATLAS:2022hbt,ATLAS:2023djh} (left panel) and the LUX-ZEPLIN spin-independent dark matter search plot extracted from Ref.~\cite{LZ:2024zvo} (right panel) for $2.25 \le m_{\widetilde{g}} \le 2.8$~TeV (top panel), which is within the HL-LHC $5 \sigma$ discovery region, and $m_{\widetilde{g}} > 2.8$~TeV (bottom panel), which is beyond the HL-LHC $5 \sigma$ discovery region. The red arrow on the $m_{\widetilde{g}} > 2.8$~TeV LUX-ZEPLIN plot indicates that the parameter space continues to the right to larger $m_{\widetilde{\chi}_1^0}$, which is beyond the scanning range in this work. However, all these points with $m_{\widetilde{\chi}_1^0} \gtrsim $ 1~TeV belong to spectra with $m_{\widetilde{g}} \gtrsim$ 3.1~TeV, which is beyond the HL-LHC $5 \sigma$ discovery region. Note though that the spin-independent cross-section $\sigma_{SI}$ does not decrease beyond the minimum shown in the plots. The gap in the red points on the right-side panel emerges from the ATLAS electroweak production of charginos and sleptons decaying into dilepton final states~\cite{ATLAS:2019lff}, represented by the gold colored exclusion region in the left-side panel ATLAS summary plot.}
    \label{atlas_luxzeplin}
\end{figure*}

{\bf Numerical Procedure~--}~The numerical calculations here follow the same process as Ref.~\cite{Yin:2023jhs}, namely we concentrate on small $M_{1}$ amidst a wide scan of the $M_5$ and $M_{1X}$ gaugino mass scales. This necessitates a committment to the $\mathcal{F}$-$SU(5)$ relationship amongst $M_{1}$ for $U(1)_{Y}$ and $M_{1X}$ for $U(1)_{X}$ at the $M_{32}$ scale, which is~\cite{Li:2010rz}:
\begin{equation}
\frac{M_1}{\alpha_1} \equiv \frac{24}{25} \frac{M_{1 X}}{\alpha_{1 X}}+\frac{1}{25} \frac{M_5}{\alpha_5}~,
\end{equation}
where $\alpha_i$ are the gauge couplings at their respective scales. In tandem, we require compliance with these phenomenological constraints:
\begin{itemize}
    \item Neutralino LSP
    \item Constraints on the mass of gluino and first and second generation squark masses~\cite{ATLAS:2017mjy,Vami:2019slp,CMS:2017okm} of 
    $m_{\widetilde{g}}\gtrsim 2.25 ~\text{TeV} $,~$m_{\widetilde{q}}\gtrsim 2.0 ~\text{TeV}$
    \item Rare B-meson decay constraint of~\cite{CMS:2014xfa} $1.6 \times 10^{-9}$ $\leq$ BR($B_{s}^0 \rightarrow \mu^{+}\mu^{-}$) $\leq$ $4.2 \times 10^{-9}$ and branching ratio of rare b-quark decay of $2.99 \times 10^{-4} \leq$ BR($b\rightarrow s \gamma$) $\leq 3.87 \times 10^{-4}$~\cite{HFLAV:2014fzu}
    \item Adherence to the measured light Higgs boson mass and its experimental uncertainty plus a small theoretical uncertainty~\cite{ATLAS:2012yve,CMS:2012qbp,Slavich:2020zjv,Allanach:2004rh}, giving a range of 123 GeV$\leq m_{h} \leq$ 127 GeV
    \item Constraints on spin-independent DM-nuclei cross sections from XENONnT~\cite{XENON:2023cxc} and LUX-ZEPLIN~\cite{LZ:2018qzl,LZ:2022lsv}
    \item 5$\sigma$ Planck 2018~\cite{Planck:2018nkj} cold dark matter relic density observations of $0.114 \leq \Omega_{\rm DM}h^{2} \leq 0.126$.
\end{itemize}

All numerical computations are carried out with a version of MicrOMEGAs 2.1~\cite{Belanger:2001fz} that integrates proprietary code modifications within the embedded SuSpect 2.34~\cite{Djouadi:2002ze} codebase necessary to include vector-like particle contributions to RGEs that are further enhanced to also incorporate flipped $SU(5)$. Annihilation and coannihilation relic density calculations $\Omega_{\chi} h^2$ and spin-independent scattering cross-sections $\sigma_{SI}$ are undertaken in MicrOMEGAs, as are rare decay processes BR($B_{s}^0 \rightarrow \mu^{+}\mu^{-}$),  BR($b\rightarrow s \gamma$), and $\Delta a_{\mu}$ of the $(g_{\mu}-2)$ of the muon experiments. Furthermore, we choose the sign of the SUSY breaking parameter $\mu$ to be positive for all points, $i.e.$ sgn$(\mu) > 0$, as suggested by results of $(g_{\mu}-2)$.

The light Higgs boson mass was observed to be $m_h = 125.1 \pm 0.14$~ GeV~\cite{ATLAS:2012yve,CMS:2012qbp}, though we grant tolerance for a 2$\sigma$ experimental uncertainty beyond a small theoretical uncertainty, giving an effective range of 123 GeV$\leq m_{h} \leq$ 127 GeV. The vector-like flippons and Higgs fields will couple by virtue of the flippon Yukawa coupling, although, the flippon contribution to $m_h$ and the vector-like decoupling scale $M_V$ are inversely proportional~\cite{Huo:2011zt}. For $M_V \sim 10$~TeV, the flippon contribution through 1-loop can can still contribute up to $\sim$2.5~GeV to $m_h$, with contributions of $\sim$17~GeV to $m_h$ typical for $M_V \sim 2$~TeV. To reach the light Higgs boson measured value in our analysis here, all points include the maximal flippon Yukawa coupling, in addition to the 2-loop coupling to the heavy light stops via the top Yukawa coupling.

The primary factor in identifying the bulk for a pure bino-like neutralino involves avoiding resonance annihilation and coannihilation. The lightest neutralino $\widetilde{\chi}_1^0$ is a viable DM candidate given its eigenstate can be formulated as a linear combination of the bino ($\widetilde{B}$), wino ($\widetilde{W}$), and two neutral Higgsinos ($\widetilde{H}^0_1, \widetilde{H}^0_2$)~\cite{Ellis:1983ew}
\begin{equation}
\tilde{\chi}_1^0=N_{11} \tilde{B}+N_{12} \tilde{W}^3+N_{13} \tilde{H}_1^0+N_{14} \tilde{H}_2^0~.
\label{DMeq}
\end{equation}
To this end the following conditions are required:
\begin{itemize}
    \item Only a minimum of 99.9\% bino-like LSPs are allowed to prevent Higgsino or wino component large annihilation cross sections, $i.e.$ $N_{11} \simeq 1$
    \item $2 m_{\widetilde{\chi}_{1}^{0}} \ll m_{H^{0}}$,$m_{A^{0}}$ and $2 m_{\widetilde{\chi}_{1}^{0}} \gg m_{h}$ are enforced to inhibit the "Higgs funnel"
    \item First and second right-handed sleptons, $\widetilde{\tau}_{1}$ and $\widetilde{e}_{R}$, are naturally light, so coannihilation processes are negligible when $\mathcal{R}_{\widetilde{\tau}_{1}}\equiv \frac{m_{\widetilde{\tau}_{1}}-m_{\widetilde{\chi}_{1}^{0}}}{m_{\widetilde{\chi}_{1}^{0}}} \gtrsim 10 \%$ and $\mathcal{R}_{\widetilde{e}_{R}}\equiv \frac{m_{\widetilde{e}_{R}}-m_{\widetilde{\chi}_{1}^{0}}}{m_{\widetilde{\chi}_{1}^{0}}} \gtrsim 10 \%$
\item SM Higgs resonance must vanish, which occurs when $|\mu|^{2} \gg M_{Z}^{2}$, via the coupling $g_{h{\widetilde{\chi}_{1}^{0}}{\widetilde{\chi}_{1}^{0}}} \propto \frac{M_{Z}(2\mu \text{cos}\beta+ M_{1})}{\mu^{2}-M_{1}^{2}}$~\cite{Djouadi:2005dz}.
\end{itemize}

\begin{figure*}
    \centering
    \begin{tabular}{c c}
        \includegraphics[width=1\textwidth, trim = {0 10cm 0 0}, clip]{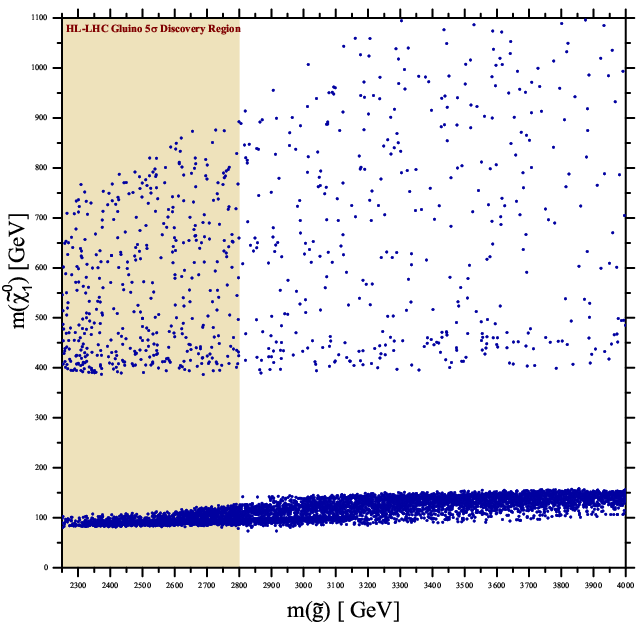} & \hspace*{-10.0cm}\includegraphics[width=1\textwidth, trim = {0 10cm 0 0}, clip] {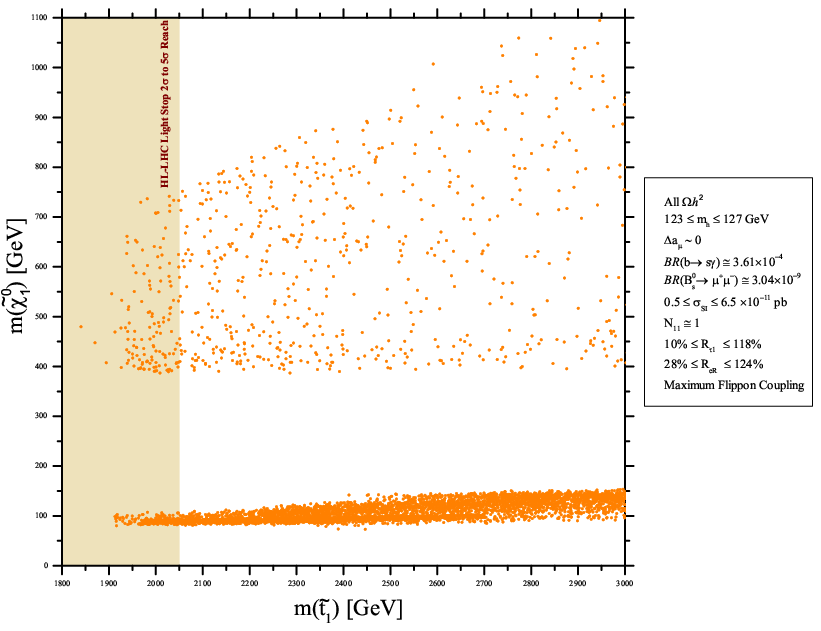} \\
        \includegraphics[width=1\textwidth, trim = {0 10cm 0 0}, clip]{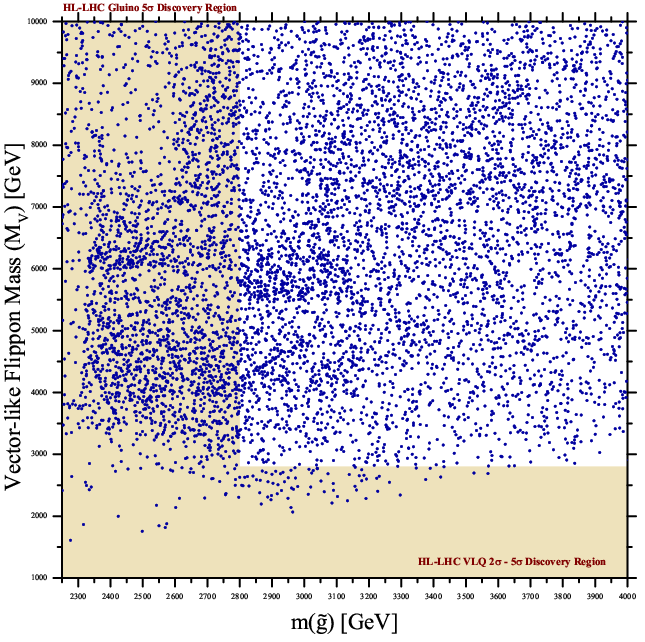} & \hspace*{-10.0cm}\includegraphics[width=1\textwidth, trim = {0 10cm 0 0}, clip] {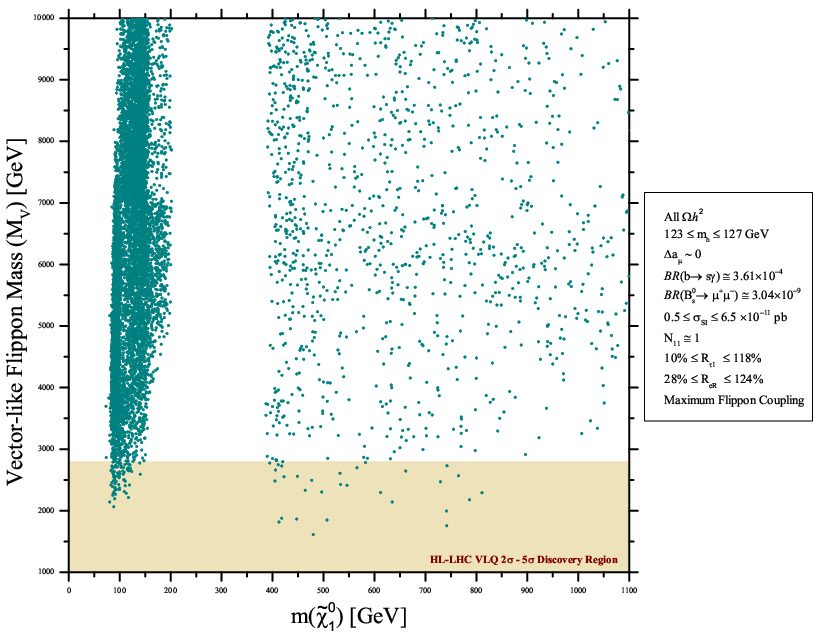}
     \end{tabular}
    \caption{Gluino versus LSP (top left) and light stop versus LSP (top right). Gluino versus vector-like flippon mass $M_V$ (bottom left) and LSP versus vector-like flippon mass $M_V$ (bottom right). The shaded rectangles on the left side of the panels on the left-hand side of this FIG.~\ref{atlas_luxzeplin_gt_2800GeV} identify the gluino HL-LHC 5$\sigma$ discovery zone, applied as $2.25 \leq m_{\widetilde{g}} \leq 2.8$~TeV. The shaded rectangle on the left side of the upper right panel signifies the light stop HL-LHC 2$\sigma$ to 5$\sigma$ reach of about $m_{\widetilde{t}_1} \lesssim 2.0 - 2.05$~TeV. The shaded rectangle at the bottom of the bottom half panels designates the VLQ discovery region, adopting the maximum projected upper VLQ limit of 2.8~TeV.}
    \label{atlas_luxzeplin_gt_2800GeV}
\end{figure*}

{\bf Numerical Results~--}~The principal motivation is an immediate refresh of the generalized No-Scale ${\cal F}$-$SU(5)$ status with respect to the LUX-ZEPLIN 1000-day experiment. The present situation is depicted in FIG.~\ref{atlas_luxzeplin}, where the right-side panel highlights the ongoing probe by LUX-ZEPLIN into the model space. Red points graphed in FIG.~\ref{atlas_luxzeplin} represent the generalized No-Scale ${\cal F}$-$SU(5)$ spin-independent scattering cross-sections computed as prescribed in the prior section, superimposed upon the LUX-ZEPLIN spin-independent dark matter search plot extracted from Ref.~\cite{LZ:2024zvo}. The projected sensitivity of the HL-LHC to a 5$\sigma$ gluino discovery is about $m_{\widetilde{g}} \lesssim 2.8$~TeV~\cite{Baer:2025zqt,ATLAS:2025eii,Baer:2026zra}, therefore, the right-side panels of FIG.~\ref{atlas_luxzeplin} are segregated into $2.25  \leq m_{\widetilde{g}}  \leq 2.8$~TeV (upper) and $m_{\widetilde{g}} > 2.8$~TeV (lower). An intriguing aspect of the generalized No-Scale ${\cal F}$-$SU(5)$ model space depicted in FIG.~\ref{atlas_luxzeplin} is illuminated by the LUX-ZEPLIN 1000-day experiment ``Median 3$\sigma$ discovery potential'' demarcation line distinctly bisecting the bulk, indicating $the ~bulk ~is ~imminently ~under ~test ~at ~this ~moment$. The generalized No-Scale ${\cal F}$-$SU(5)$ parameter space is wholly contained within the upper right $2.25  \leq m_{\widetilde{g}}  \leq 2.8$~TeV panel and precisely delineated by the red points due to the 2.8~TeV gluino cutoff, which correlates to a $\widetilde{\chi}_{1}^{0}$ upper limit of $m_{\widetilde{\chi}_{1}^{0}} \leq 960$~GeV. Note though that for $0.114 \leq \Omega_{\chi}h^{2} \leq 0.126$, the points are very tightly constrained to $86 < m_{\widetilde{\chi}_{1}^{0}} <103$~GeV. However, when studying the more massive gluinos beyond the HL-LHC 5$\sigma$ discovery upper limit of about 2.8~TeV, the LSP can persevere to indefinitely higher levels well beyond our scanning ranges applied in this analysis. Thus, the intent of the large arrow in the lower right panel of FIG.~\ref{atlas_luxzeplin}, indicating the perpetual extension of the generalized No-Scale ${\cal F}$-$SU(5)$ points to larger $m_{\widetilde{\chi}_{1}^{0}}$.

Before continuing our discussion of numerical results, we pause to offer some salient details regarding spin-independent scattering cross-sections. Spin-independent nuclear recoil from elastic collisions between nucleons and bino LSPs in the bulk are under probe by direct detection experiments such as LUX-ZEPLIN. Neutralinos couple to nuclei through the total mass of the nucleus of the target material, where primary neutralino DM couplings to nuclei are scalar-mediated interactions via squarks and Higgs bosons. The scattering cross-section is proportional to the mass-squared of the nucleus so the neutralino DM coherently couples to the entire nucleus. Independent of spin, it therefore vastly boosts the cross-section with massive nuclei like xenon. We inspect spin-independent DM-proton scattering cross-sections given the spin-independent DM-proton and DM-neutron cross-sections diverge only negligibly by virtue of commensurate form-factors.

Any looming event observed by LUX-ZEPLIN must be confirmed by the LHC3 or HL-LHC, thus the left-side panel of FIG.~\ref{atlas_luxzeplin} shows the bulk conveniently wedged within a hard to observe soft lepton SUSY search region. The left panels of FIG.~\ref{atlas_luxzeplin} display the bulk region superimposed upon the summary plot of ATLAS SUSY searches for electroweak production of sleptons~\cite{ATLAS:2023xco,ATLAS:2014zve,ATLAS:2019lng,ATLAS:2019lff,ATLAS:2022hbt,ATLAS:2023djh}. Similarly, though not shown here, is the current status of the bulk pertaining to CMS SUSY searches for electroweak production of sleptons~\cite{CMS:2020bfa,CMS:2023qhl,CMS:2022syk}, that also persists without probe. In FIG.~\ref{atlas_luxzeplin} the superimposed green dots denote the generalized No-Scale ${\cal F}$-$SU(5)$ $\widetilde{\tau}_{1}$ and the blue dots the right-handed sleptons $\widetilde{e}_{R}$, $\widetilde{\mu}_{R}$ where $m_{\widetilde{e}_{R}} = m_{\widetilde{\mu}_{R}}$. Of note is the fact that the ATLAS orange region in Fig.~\ref{atlas_luxzeplin} represents constraints on $\widetilde{e}_{R} = \widetilde{\mu}_{R}$ rather than $\widetilde{\tau}_{1}$, therefore, there is no issue with the green $\widetilde{\tau}_{1}$ points overlaid onto the ATLAS excluded space for right-handed sleptons. The very apparent gap in the generalized No-Scale ${\cal F}$-$SU(5)$ red points in the right-side panel LUX-ZEPLIN scattering cross-section plot within FIG.~\ref{atlas_luxzeplin} arises from the ATLAS electroweak production of charginos and sleptons decaying into dilepton final states~\cite{ATLAS:2019lff}, characterized by the gold colored exclusion region in the ATLAS summary plot. This ATLAS dilepton search excludes all $\widetilde{e}_{R}$, $\widetilde{\mu}_{R}$ from $189 < m_{\widetilde{e}_{R}/\widetilde{\mu}_{R}} <633$~GeV for $2.25  \leq m_{\widetilde{g}}  \leq 2.8$~TeV and $266 < m_{\widetilde{e}_{R}/\widetilde{\mu}_{R}} <600$~GeV for $m_{\widetilde{g}} > 2.8$~TeV. Evidently from Fig.~\ref{atlas_luxzeplin}, the bulk has not yet been fully reached by ATLAS and CMS. While ATLAS has confirmed that it is still possible the impending data that completes the LHC3 phase could tease entry into the bulk, a more extensive intrusion awaits at the HL-LHC, together with the succeeding FCC-ee and CEPC future circular colliders.

Specifically with respect to the gluino, the upper left panel of FIG.~\ref{atlas_luxzeplin_gt_2800GeV} illustrates a more complete accounting, harboring the aforementioned exclusion void from ATLAS dilepton SUSY searches. The shaded rectangle on the left-side of the plot space signifies those gluinos within the HL-LHC 5$\sigma$ discovery zone. Likewise, light stops $\widetilde{t}_1$ are projected to realize a 5$\sigma$ discovery threshold at the HL-LHC of about $m_{\widetilde{t}_1} \lesssim 1.7 - 1.8$~TeV~\cite{Baer:2025zqt,ATLAS:2025eii,Baer:2026zra,Dutta:2023qxd} with an accompanying 2$\sigma$ reach of around $m_{\widetilde{t}_1} \lesssim 2.0 - 2.05$~TeV~\cite{Baer:2025zqt,ATLAS:2025eii,Baer:2026zra,Dutta:2023qxd}, explicitly demonstrated in the upper right panel of FIG.~\ref{atlas_luxzeplin_gt_2800GeV}. 

The string derived TeV-scale extra $10+\overline{10}$ vector-like multiplet and charged vector-like singlet $(XE,XE^c)$ decouple at the scale $M_V$, perhaps yielding observation as large as $M_V \lesssim 2.4$~TeV at 5$\sigma$ and $M_V \lesssim 2.8$~TeV at 2$\sigma$ for vector-like quark (VLQ) searches at the HL-LHC~\cite{Cetinkaya:2020yjf,Alves:2023ufm,Canbay:2023vmj,Sagheer:2026kad}. Current ATLAS and CMS VLQ search limits at the LHC are as low as $M_V \lesssim 1.0$~TeV but as high as $M_V \lesssim 2.1$~TeV~\cite{ATLAS:2018tnt,ATLAS:2018ziw,ATLAS:2022ozf,ATLAS:2024gyc,CMS:2024bni,ATLAS:2024xdc,ATLAS:2026ojf}, dependent upon search methodology and final states. The lower panels of FIG.~\ref{atlas_luxzeplin_gt_2800GeV} reveal the vector-like flippon mass scale $M_V$ with respect to the gluino and LSP. The horizontal shaded rectangle at the lower end of the plot space exhibits the VLQ discovery region, adopting the maximum projected upper VLQ limit of 2.8~TeV. Those generalized No-Scale ${\cal F}$-$SU(5)$ points with $M_V < 2.8$~TeV solely reside external to the 5$\sigma$ Planck 2018 cold dark matter relic density observations of $0.114 \leq \Omega_{\rm DM}h^{2} \leq 0.126$ if $m_{\widetilde{g}} \leq 2.8$~TeV. For points satisfying the 5$\sigma$ Planck 2018 measurements and $m_{\widetilde{g}} \leq 2.8$~TeV, the vector-like flippon mass scale is $M_V \gtrsim 4$~TeV. For $m_{\widetilde{g}} > 2.8$~TeV and $0.114 \leq \Omega_{\chi}h^{2} \leq 0.126$, the vector-like flippon mass scale can be as low as $M_V \gtrsim 2.8$~TeV, right at the threshhold of conceivable HL-LHC observation.

Fifteen benchmark points are enumerated in TABLE~\ref{table} partitioned into three groups bounded by undersaturated, saturated, and oversaturated DM relic density. Considering that our primary effort here is to submit testable predictions for the LUX-ZEPLIN 1000-day experiment and the HL-LHC, the benchmarks of TABLE~\ref{table} are collectively $m_{\widetilde{g}} \leq 2.8$~TeV, and thus within the HL-LHC 5$\sigma$ discovery region. Notice that the rare decay processes are not tabulated in TABLE~\ref{table} due to invariant results for the full generalized No-Scale ${\cal F}$-$SU(5)$ model. Calculations yield uniform numerical values of BR$(B_{s}^0 \rightarrow \mu^{+}\mu^{-}) \simeq 3.04 \times 10^{-9}$,  BR$(b\rightarrow s \gamma) \simeq 3.61 \times 10^{-4}$, and $\Delta a_{\mu} \sim 0$. It has been well demonstrated in the plethora of literature over the prior 16 years that No-Scale ${\cal F}$-$SU(5)$ produces negligible SUSY contributions to $(g_{\mu}-2)$ of the muon~\cite{Li:2010ws}, which temporarily stood in contrast to the initially reported large $\Delta a_{\mu}$ in 2021 by the Muon g-2 Collaboration~\cite{Muong-2:2021ojo}. In fact, $(g_{\mu}-2)$ of the muon persisted as the lone experiment inconsistent with No-Scale ${\cal F}$-$SU(5)$. Curiously, new hadronic contribution computational techniques have shifted the theoretical prediction close to the empircally measured value, furnishing $\Delta a_{\mu} = 38(63) \times 10^{-11}$~\cite{Aliberti:2025beg}, conclusively erasing the tension. More precisely, all points in generalized No-Scale ${\cal F}$-$SU(5)$ produce numerical values of $-10 \times 10^{-11} \lesssim \Delta a_{\mu} \lesssim 28 \times 10^{-11}$, such that the revised $\Delta a_{\mu}$ of Ref.~\cite{Aliberti:2025beg} is now compatible with No-Scale ${\cal F}$-$SU(5)$.

{\bf Conclusion~--}~High-energy theory has been relentlessly advancing for several decades, but in recent years attention has shifted to measurement and observation as technology is finally approaching par with theories awaiting test. Two present and forthcoming materializations of this are the LUX-ZEPLIN 1000-day experiment and the high luminosity LHC (HL-LHC), both projected to probe to unprecedented scales and reach energies the high-energy physics community only dreamt of just a mere 20 years ago. In light of these long anticipated high-tech breakthroughs applied to scientific discovery, we toss our hat into the ring with the goal of uncovering the physical manifestation of dark matter and natural emergence of supersymmetry (SUSY) within string inspired grand unification theories (GUTs). 

The GUT we study here, No-Scale ${\cal F}$-$SU(5)$, has been on its own journey since flipped $SU(5)$ was initially concieved in 1982, with numerous successive iterative advancements along the way, leading to its phenomenological development in the previous 20 years that finally supported its long anticipated opportunity to be put to the test. This long history of incremental development has sharpened the model to a fine point such that it is consistent with all currently operating experiments in the long voyage of discovery for supersymmetry, dark matter, and a Theory of Everything. Make no mistake, No-Scale ${\cal F}$-$SU(5)$ is not one of the ``johnny-come-lately''~\cite{briggs} GUT models just now debuting on the scene, but a model with decades of cumulative progress leading to its elegant contemporary construction that we study here. 

The present-day version is generalized No-Scale ${\cal F}$-$SU(5)$ and inherently includes naturalness by means of Effective Super-Natural SUSY, solving the electroweak fine-tuning (EWFT) problem. The generalization of Effective Super-Natural SUSY extends the naturalness objective by extracting those regions of the parameter space supporting only annihilation for dark matter, and not the more finely tuned scenarios known as the $Z$/Higgs funnel or $Z$/Higgs resonance, coannihilation, and large Higgsino or wino components. This offers a more comprehensive package of naturalness than just attention to EWFT. By considering annihilation only, a light LSP bino dark matter and light right-handed sleptons can coexist with very heavy gluino and squark masses, substantial enough to eclipse LHC Run 3 (LHC3) constraints on gluinos and squarks. Moreover, the Planck 5$\sigma$ dark matter relic density measurements can be readily satisfied by the light bino mass encountering annihilation, without resorting to one of the more finely tuned scenarios. Light bino LSPs and light right-handed sleptons embody what was referred to as the bulk, a popular term expressed in the early days of supersymmetry, in what could be regarded as a 1980s reunion with flipped $SU(5)$ here in 2026.

The work here demonstrates that the bulk in generalized No-Scale ${\cal F}$-$SU(5)$ is immediately under test by the LUX-ZEPLIN 1000-day experiment, thus any prospective observations of spin-independent dark matter-nucleon scattering events in 2026-27 will require confirmation by the HL-LHC once the newest LHC configuration powers up in 2030. The bulk dwells in the rather soft lepton region that has been challenging thus far for the LHC to probe, though discussions with the ATLAS Collaboration indicate their members are confident the bulk could be penetrated within the forthcoming LHC3 data and subsequent HL-LHC. Just as promising though are the associated heavy gluinos and light stops that certainly will be within reach of the HL-LHC, up to $m_{\widetilde{g}} \lesssim 2.8$~TeV at 5$\sigma$ and $m_{\widetilde{t}_1} \lesssim 2.05$~TeV at 2$\sigma$. Meanwhile, the string derived TeV-scale extra $10+\overline{10}$ vector-like multiplet and charged vector-like singlet $(XE,XE^c)$, referred to as flippons, are modelled numerically by the decoupling scale $M_V$, the scale at which all vector-like particles decouple from the RGEs. This decoupling scale could also be within reach of HL-LHC vector-like quark searches, possibly as high as $M_V \lesssim 2.8$ TeV, at least for a partition of the generalized No-Scale ${\cal F}$-$SU(5)$ parameter space that generates undersaturated and oversaturated dark matter relic density. The vector-like flippons for saturated relic density are probably beyond the HL-LHC reach given those points compel $M_V \gtrsim 4$~TeV.

The LHC era will long be commemorated for the observed light Higgs boson, but the machine may yet deliver another milestone(s) to supplement its legacy. Likewise, the dark matter direct detection experiments have endured some tantalizing statistics, though none to date have survived meticulous scrutiny. With steadily advancing technology devoted to scientific exploration and progressively improving machine learning algorithms aiding data search tactics, we may yet find our answer to the Theory of Everything in the coming decade.

\begin{acknowledgments}
     We thank the ChinaHPC for providing us with the computational platform and Xiangwei Yin for providing the computational data. This research is supported in part by the National Key Research and Development Program of China Grant No. 2020YFC2201504, by the Projects No. 11875062, No. 11947302, No. 12047503, No. 12275333, and No. 12347101 supported by the National Natural Science Foundation of China, by the Key Research Program of the Chinese Academy of Sciences, Grant No. XDPB15, by the Scientific Instrument Developing Project of the Chinese Academy of Sciences, Grant No. YJKYYQ20190049, by the International Partnership Program of Chinese Academy of Sciences for Grand Challenges, Grant No. 112311KYSB20210012.
\end{acknowledgments}


\end{document}